*Article*

# Thermal Infrared Imaging to Evaluate Emotional Competences in Nursing Students: A First Approach through a Case Study

Pilar Marqués-Sánchez [1], Cristina Liébana-Presa [1,*], José Alberto Benítez-Andrades [2], Raquel Gundín-Gallego [3], Lorena Álvarez-Barrio [4] and Pablo Rodríguez-Gonzálvez [5]

1. SALBIS Research Group, Faculty of Health Sciences, Campus of Ponferrada, University of León, 24401 Ponferrada, Spain; pilar.marques@unileon.es
2. SALBIS Research Group, Department of Electric, Systems and Automatics Engineering, University of León, 24071 León, Spain; jbena@unileon.es
3. Bierzo Hospital, 24401 Ponferrada, Spain; rgundg@gmail.com
4. Department of Nursing and Physiotherapy, Faculty of Health Sciences, Campus of Ponferrada, University of León, 24401 Ponferrada, Spain; lalvb@unileon.es
5. Department of Mining, Surveying and Structure, Campus of Ponferrada, University of León, 24401 Ponferrada, Spain; p.rodriguez@unileon.es
* Correspondence: cliep@unileon.es

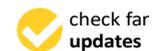

**Abstract:** During university studies of nursing, it is important to develop emotional skills for their impact on academic performance and the quality of patient care. Thermography is a technology that could be applied during nursing training to evaluate emotional skills. The objective is to evaluate the effect of thermography as the tool for monitoring and improving emotional skills in student nurses through a case study. The student was subjected to different emotions. The stimuli applied were video and music. The process consisted of measuring the facial temperatures during each emotion and stimulus in three phases: acclimatization, stimulus, and response. Thermographic data acquisition was performed with an FLIR E6 camera. The analysis was complemented with the environmental data (temperature and humidity). With the video stimulus, the start and final forehead temperature from testing phases, showed a different behavior between the positive (joy: 34.5 °C–34.5 °C) and negative (anger: 36.1 °C–35.1 °C) emotions during the acclimatization phase, different from the increase experienced in the stimulus (joy: 34.7 °C–35.0 °C and anger: 35.0 °C–35.0 °C) and response phases (joy: 35.0 °C–35.0 °C and anger: 34.8 °C–35.0 °C). With the music stimulus, the emotions showed different patterns in each phase (joy: 34.2 °C–33.9 °C–33.4 °C and anger: 33.8 °C–33.4 °C–33.8 °C). Whenever the subject is exposed to a stimulus, there is a thermal bodily response. All of the facial areas follow a common thermal pattern in response to the stimulus, with the exception of the nose. Thermography is a technique suitable for the stimulation practices in emotional skills, given that it is non-invasive, it is quantifiable, and easy to access.

**Keywords:** emotions; educational innovation; nursing students; thermography; case study

## 1. Introduction

The emotional skills of nursing professionals are key in the carrying out of their professional work [1]. During their period of university studies, it is important to develop and hone emotional skills for their impact on academic performance and the quality of patient care [2]. However, emotional skills are not usually included in the nursing syllabus even though it is a useful aspect when faced with the complexity of the health care environment [3]. Codie et al. [4] highlight that those nurses do have the capacity to identify their emotions but often do not have the habit of using the skill of Emocional Competence. The training for developing skills are in line with the current context of educational reform in which the focus is now on competency-driven approaches to instruction, and not so much on the processes based on memorizing [5]. In spite of the lack of studies that apply methods to measure and monitor emotions, it has been detected that thermography is a technology which could be applied during nursing training in order to evaluate and train emotional skills.

The basic principle of thermography is that the variation in temperature of an element can be identified from the energy radiated by means of the electromagnetic spectrum. These data can be gathered by means of non-invasive techniques such as infrared thermographic cameras [6]. A thermographic camera provides a thermal image, or thermogram, of the superficial distribution of body temperature. Since the skin temperature is related to the blood circulation, it can be related to physiological parameters, such as emotions [7].

The superficial body temperature is computed by means of the Planck's law. In the case of the skin, its emissivity value is difficult to measure accurately, but previous studies agree that the emissivity of an intact skin is approximately $0.96 \pm 0.03$ [8]. When it is analyzed, the human body temperature, the subject has to be still, and the subject should also not ingest alcohol and smoke prior the test, since these actions can cause thermal variations [9]. In addition, the data acquisition should be carried out without any direct light source or even in dark. Thermographic images must be applied carefully, and not limited to the simple visualization. While it is true that the application of a color palette improves the comprehension of the intrinsic information provided in the thermal imaging, the comparison among the results is constrained by the color scale, which could be linear, logarithmic, or even non-standard palettes [10]. Therefore, to avoid this subjectivity, the analysis has to be carried out in term of raw thermal values. In addition, the areas for thermal evaluation have to be standardized to a proper comparison and evaluation. In the case of the face, the typical Regions Of Interest (ROI) are mouth, nasal tip, forehead, cheeks, and eyes [11].

There is a large amount of evidence for the application of thermography in health; for example, the physiological response of infants when they are in a joyful emotion [7], the affective responses in social interaction [12], the identification of the skin temperature behavior after training of lower-limbs [13], and the analysis of stress in young people [14]. Furthermore, there is large number of research works related to the application of thermography to study the human emotional and affective state [15]. Emotions are present in everyday life and, if nursing students are not prepared to face the challenging emotional work of care, they may feel vulnerable [16].

The teaching interest in thermography is its non-invasive character, its capacity to gather data on emotions which are difficult to quantify using other therapeutic techniques and the low cost of the device. Looking at all of its aspects, the question is: what would be the best way of incorporating thermography into the emotional education process of nursing students? One strategy would be to incorporate thermography into the simulation practices for training emotional skills. Simulation is a method which is becoming more and more used with undergraduate nursing students in contexts in which emotions are present, for example, to train behavior with patients at the end-of-life [17], for the measurement of cognitive load and emotional impact in an ICU environment [18] and for the training of cultural sensitivity [19], among others. It is useful in measuring the progressive assessment and competency evaluation [20], for the application of learner-centered teaching methods in nursing education [21].

On the other hand, this proposal would permit the students to access the specialized practices and materials in an asynchronous way thanks to advances in ICT in the educational panorama, creating virtual laboratories. Virtual laboratories are postulated as reinforcing and supporting tools that encourage self-learning, and promote participatory environments [22], even improving the motivation [23]. The application of virtual laboratories in the field of Health Sciences is still very limited in spite of the growing range of online training programs. Virtual laboratories are not only looking to improve the acquisition of practical skills that require the use of specialized equipment, but also allow the evaluation of the degree of acquisition of the aforementioned skills on the part of the students.

Therefore, the aim of this study is to carry out a first approximation to apply thermography to the monitoring of emotions in nursing students and propose it as a teaching technique for virtual laboratories. To do so, the thermal bodily responses emitted during two emotional stimuli: Audiovisual (video) and listening (music) will be described.

This research has a double impact. On the one hand, to add evidence to the field of thermography and the emotions, and, on the other hand, to incorporate new knowledge into this first approximation, as regards thermography applied to emotional skills in the simulation practices in virtual laboratories.

The emotional competences learning is facilitated even more through the visualization of the temporal sequence images when faced with a stimulus. Therefore, we proposed the upload of the IR images, stimuli, and analyzed response to the learning management system of the institution together with the didactical package similarly as proposed by [24] and [25]. As a result (Figure 1), students could benefit from an asynchronous competence learning and evaluation. Moreover, this approach encourages autonomous learning of students.

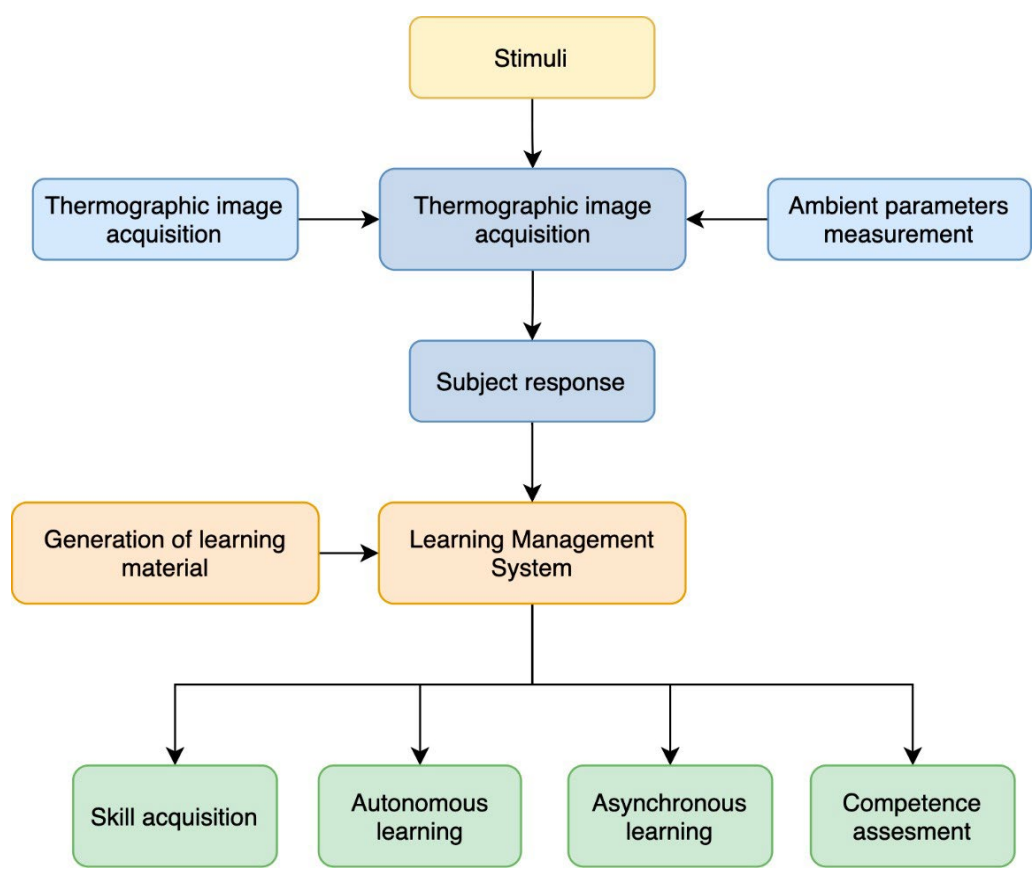

**Figure 1.** Flow graph of the proposed learning approach.

## 2. Methods

*2.1. Study Design*

A descriptive study has been developed by means of the methodology used to study the case. This methodology allows decisions to be taken on the processes and causalities to be studied. It is suitable to be able to respond to the questions about when and why, when faced with a series of events, and it is valid to apply it to research in which the theory still does not give rise to suitable responses [26]. Fox-Wolfgramm [27] underlined the use of this method in contexts in which the researchers wanted to develop a theory from the analysis of a process. For Eisenhardt [28], the method provided empirical checking and validity that arise from the intimate linkage with empirical evidence.

*2.2. Experiment Setup*

The subject was a thirty-two-year-old woman, a fourth year student in Nursing Studies (Public University, Spain). The participant signed an informed consent letter, following the protocol of previous similar studies [29].

The room for the taking of the data was chosen on the basis of the requirements mentioned by Fernandez-Cuevas et al. [30]: the minimization of direct solar radiation, the absence of guttering, pipes, and non-controlled heat pipes, a clean and manageable room with minimum sources of light. In the room, there was a table, two chairs, the thermal camera, and a portable computer to show the different emotions (video and music) the subject. The distance between the camera and the subject was 80 cm to ensure that the FOV covered the entire face of the subject, and 70 cm between the portable computer and the subject. This portable computer was placed on a table located between the subject and the camera, keeping the computer out of the camera's field of view. Figure 2 details the general methodology proposed for this study.

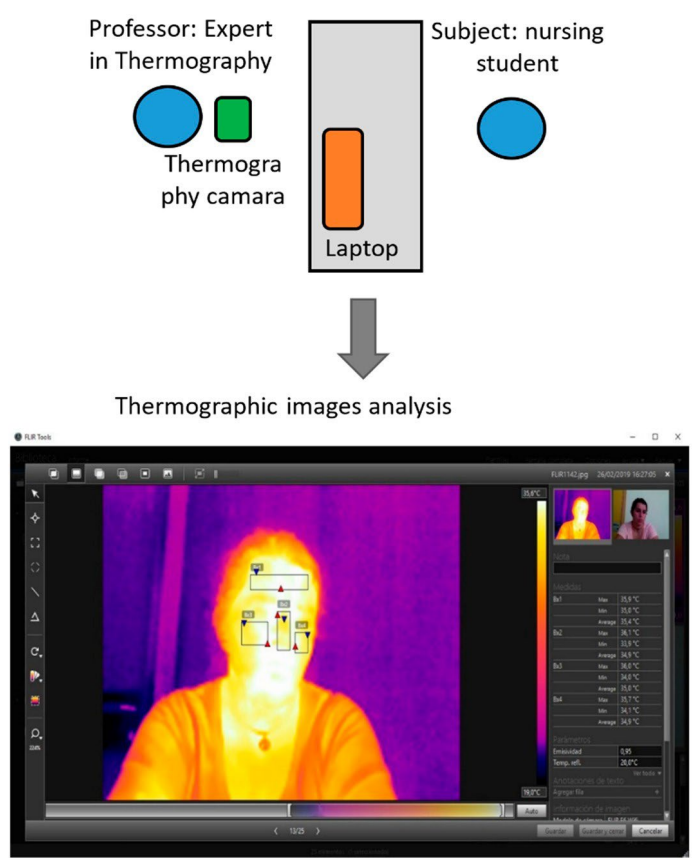

**Figure 2.** General methodology proposed for the study of emotions.

*2.3. Materials*

Thermographic data acquisition was performed with an FLIR E6 camera (FLIR Systems, Wilsonville, Oregon, United States), measuring temperatures ranging from −20 °C to 250 °C, with a thermal sensitivity of 0.06 °C, whereas the expected accuracy of temperature measurement is ±2 °C or ±2% of reading. Its sensor is a Focal Plane Array, uncooled microbolometer, with a spatial resolution of 5.2 mrad (160 × 120 pixels), capturing radiation between a wavelength of 7.5 µm and 13 µm. The minimum focus distance is 0.5 m. A Testo 605-H1 temperature measuring probe was used to measure the environmental conditions of the room. It had a precision of ±0.5 °C in temperature and ±3% relative humidity. Due to the thermal camera used, the presence of an operator was necessary to interact with the camera's manual trigger. However, since this is a first approach to evaluate emotional competences in nursing students, the thermal device could be replaced by an automatic one; therefore, the temperature measurement can be programmed as a time-lapse, and the subject could be alone in the room.

The analysis of the thermographic images was carried out using FLIR Tools software. Furthermore, a mobile telephone with a chronometer was used to establish the intervals between the capturing of the images, together with a ballpoint pen and notebook to take notes relative to the conditions of the room and/or subject.

*2.4. Data Acquisition Protocol*

The emotions chosen for carrying out the study were happiness, love, cheerfulness, anger, fear and sadness, in line with the recommendations of Nummenmaa et al. [31]. It was carried out over a period of seven days, in February and March 2019. The emotions were studied twice: the first one using audiovisual stimuli with the video, and the second one using audio stimuli by means of music [32]. As a result, twelve sessions took place.

In the first four days, the emotions brought about through the stimulus received from the video were analyzed. The chosen videos were parts of monologues, films, and shorts and a political meeting during the electoral campaign. In the last three days, emotions brought about through the music which was chosen in accordance with the musical preferences of the subject were analyzed. Two emotions were studied every day with a two-hour break between them with the exception of two days in which only one single emotion was studied. As detailed in the schedule in Table 1, the duration of each session was about thirty minutes, as suggested in specialized literature [29,30].

**Table 1.** Temporary distribution of data collection sessions (sess.).

| Video Stimulus | | | | Music Stimulus | | |
|---|---|---|---|---|---|---|
| February | | | | March | | |
| 1st sess. Joy | 2nd sess. Sadness Love | 3rd sess. Happiness Fear | 4th sess. Anger | 5th sess. Happiness Joy | 6th sess. Sadness Fear | 7th sess. Anger Love |

Taking previous studies as a base, it was pointed out to the subject that he or she had come to the sessions in compliance with a series of conditions: hair tied back, no makeup, no face cream, without having carried out physical exercise or efforts, not having consumed tea, coffee, or nicotine for at least an hour before the session [29]. During the sessions, it was recommended that they did not touch their face during the study. The subject was photographed with the thermal camera while he or she was subjected to the stimuli in such a way data were gathered on the changes in facial temperature that they experienced during the process.

The process to be followed during the entire experiment is detailed in Figure 3, following the recommendations of [15] and [29]. Each session began with an acclimatization phase of at least ten minutes (up to 15 min), during which one thermographic image was taken every minute. In this way we were able to establish the baseline of the thermal response of the subject. Immediately after that, we went on to the simulation phase, in which the student is exposed to an audio or audiovisual stimulus for a duration of between two and ten minutes. During this phase, thermographic images were taken at the beginning (pre-stimulus), at the end (the end of the stimulus), and for one minute during the experiment. At the end of the stimulus, a response phase is established in which the effect of the emotional stimulus is monitored (by means of facial temperature) and his or her tendency. For this reason, one thermal image was taken every minute.

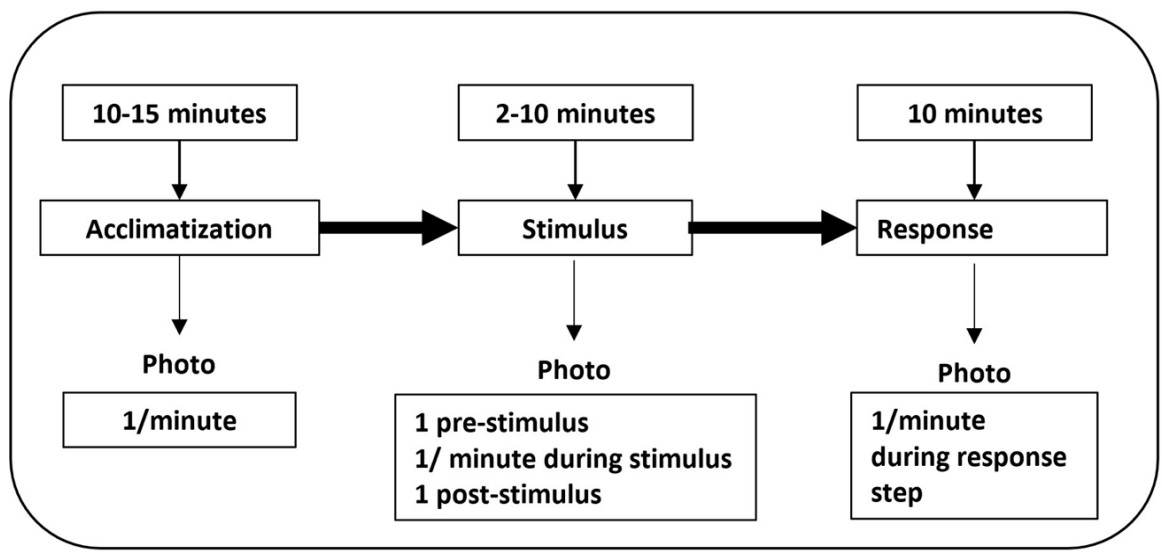

**Figure 3.** Workflow for data acquisition. During taking of thermographic images—protocol application.

In Table 2, the temperature and humidity of the room is shown at the main moments when gathering the data. These data are relevant, since, as similar studies have taken place, the environmental conditions may influence the body temperature of the subject [15].

**Table 2.** Environmental conditions of each session.

|   | Emotions | Temp/Humidity | Start Acclimatization | Start Stimulus | Final Stimulus | Final Period Response |
|---|---|---|---|---|---|---|
| **V I D E O** | Joy | Temp (°C) | 20.4 | 20.6 | 20.7 | 21.0 |
| | | Humidity (%) | 36.8 | 36.8 | 36.2 | 35.8 |
| | Sadness | Temp (°C) | 24.4 | 23.5 | 23.4 | 23.4 |
| | | Humidity (%) | 31.8 | 31.8 | 32.3 | 32.5 |
| | Love | Temp (°C) | 21.3 | 21.5 | 21.4 | 21.3 |
| | | Humidity (%) | 34.1 | 34.0 | 34.1 | 34.4 |
| | Happiness | Temp (°C) | 23.6 | 23.7 | 23.8 | 23.9 |
| | | Humidity (%) | 26.1 | 26.1 | 26.4 | 26.4 |
| | Fear | Temp (°C) | 21.9 | 21.8 | 21.7 | 21.5 |
| | | Humidity (%) | 30.0 | 31.0 | 31.5 | 32.4 |
| | Anger | Temp (°C) | 23.3 | 23.1 | 23.3 | 23.6 |
| | | Humidity (%) | 29.4 | 32.1 | 30.8 | 30.7 |
| **M U S I C** | Joy | Temp (°C) | 19.3 | 19.4 | 19.4 | 19.3 |
| | | Humidity (%) | 31.9 | 32.2 | 32.5 | 33.0 |
| | Sadness | Temp (°C) | 20.3 | 19.3 | 19.3 | 19.3 |
| | | Humidity (%) | 33.8 | 36.4 | 37.0 | 37.2 |
| | Love | Temp (°C) | 18.7 | 18.8 | 18.8 | 18.8 |
| | | Humidity (%) | 41.3 | 41.6 | 41.8 | 42.2 |
| | Happiness | Temp (°C) | 20.9 | 20.2 | 20.1 | 19.8 |
| | | Humidity (%) | 29.6 | 31.3 | 31.6 | 32.3 |
| | Fear | Temp (°C) | 19.4 | 19.4 | 19.5 | 19.4 |
| | | Humidity (%) | 38.4 | 39.2 | 39.4 | 40.1 |
| | Anger | Temp (°C) | 18.7 | 18.8 | 18.7 | 18.7 |
| | | Humidity (%) | 38.4 | 39.7 | 39.3 | 40.1 |

In Table 3, the day on which each emotion was studied is specified, as was the duration of the stimulus and the total number of photographs taken during each stimulus.

**Table 3.** Date, duration, and photos taken in each stimulus.

| | Date | Emotion | Number of Images Per Stimulus | Total | Stimulus Duration |
|---|---|---|---|---|---|
| February | 02.21.19 | Joy | 9 | 33 | 8'32" |
| | 02.26.19 | Sadness | 4 | 24 | 2'39'' |
| | 02.26.19 | Love | 7 | 27 | 5'23'' |
| | 02.27.19 | Happiness | 11 | 31 | 9'35'' |
| | 02.27.19 | Fear | 8 | 28 | 6'08'' |
| | 02.28.19 | Anger | 8 | 28 | 6'48'' |
| March | 03.04.19 | Joy | 6 | 26 | 6'08'' |
| | 03.04.19 | Happiness | 8 | 28 | 4'17'' |
| | 03.05.19 | Sadness | 5 | 25 | 3'21'' |
| | 03.05.19 | Fear | 5 | 25 | 3'35'' |
| | 03.06.19 | Anger | 6 | 26 | 4'45'' |
| | 03.06.19 | Love | 6 | 26 | 4'39'' |

*2.5. Thermographic Image Analysis*

FLIR Tools software was used to carry out the analysis of the images during the experiment. It allowed the maximum, minimum, and average temperature to be taken through the choice of the facial areas chosen in accordance with the recommendations of [15]: forehead, nose, and left and right cheeks (Figure 4).

The variables that are the object of this study in this experiment were:

- Temperature and humidity in the room. Although during the sessions they remained stable, there were slight differences between the sessions.
- Emotions: happiness, love, cheerfulness, anger, fear, and sadness [31].
- Gender and age of subject. Although this research uses the methodology of the case of a single subject, it has been decided to gather these data to be able to replicate the process of the study in future research.

The facial areas to be measured thermographically (Bx1, Bx2, Bx3 y Bx4) were chosen on the basis of the study carried out by [29], as can be seen in Figure 4.

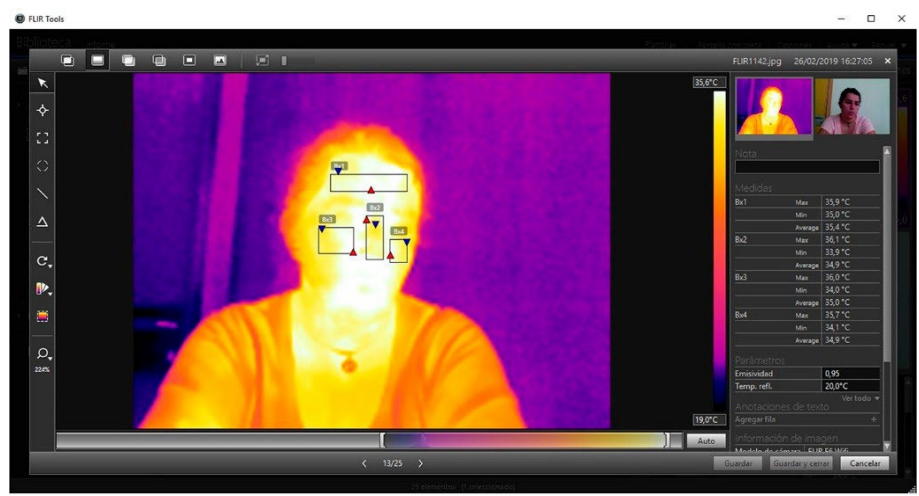

**Figure 4.** Individual image obtained during the sessions where it is possible to observe the facial areas measured with the temperature values calculated by the software.

## 3. Results

The objective of this study is to carry out a first approximation to apply thermography to the monitoring of emotions and promote it as a teaching technique for virtual laboratories. For this reason, the following has been posed: to determine the thermal bodily responses emitted when faced with different emotional stimuli and to evaluate thermography as a method applied to emotional skills by means of sequential visualization.

The results obtained are divided into three sub-epigraphs:

A. Thermal changes in the three phases, with audiovisual stimuli (video).
B. Thermal changes in the three phases, with audio stimuli (music).
C. An example of sequential thermographic evaluation.

Throughout the entire study, the same protocol was followed, evaluating the emotions and gathering the facial temperature in four regions: forehead, nose, left and right cheeks.

### 3.1. Part A: Thermal Changes with an Audiovisual Stimulation (Video)

In the acclimatization phase, it was observed that the temperature went up and down during positive emotions (joy, love, and happiness), whilst the temperature went down during negative emotions (anger and fear). Sadness showed a tendency similar to positive emotions. The temperature of testing room oscillated between 18.7 °C and 24.4 °C in the different sessions, and it was not possible to keep it constant, as had been stated.

During the stimulus phase, findings show an increase in facial temperature in all of the emotions, with a slight reduction at 3 min in anger and 8 min in happiness.

During the response phase, it was observed that there was always an increase in temperature at the beginning. This increase was maintained in all of the emotions except in sadness and love, which showed a constant decline (Figure 5). Figure 5 details the facial temperatures in the acclimatization, stimulus, and response phases. In the upper part, the emotion and the date of taking the thermographic images are identified. The graph captures the average temperatures of: forehead, nose, and right and left skin on cheeks. In the lower part, the three phases of the protocol and the minutes in which photographs were taken are shown. As confirmed, no large thermographic differences in the facial areas are perceived. Perhaps the part in which the temperature varies the most is the nasal area.

The emotions that show more variable patterns were Joy and Happiness, more specifically in the stimulation phase. In Joy, facial measurements increased especially in the area of the nose, possibly because the nose is an extreme zone in which the temperature fluctuates more than in the rest of the facial areas. In Happiness, the nasal temperature went down brusquely.

In Table 4, the thermal changes in the facial areas chosen in the study during the initial and final phase of each study are observed (acclimatization, video, and response), registered in the subject whilst being subjected to the audiovisual stimulus (video) of the chosen emotions.

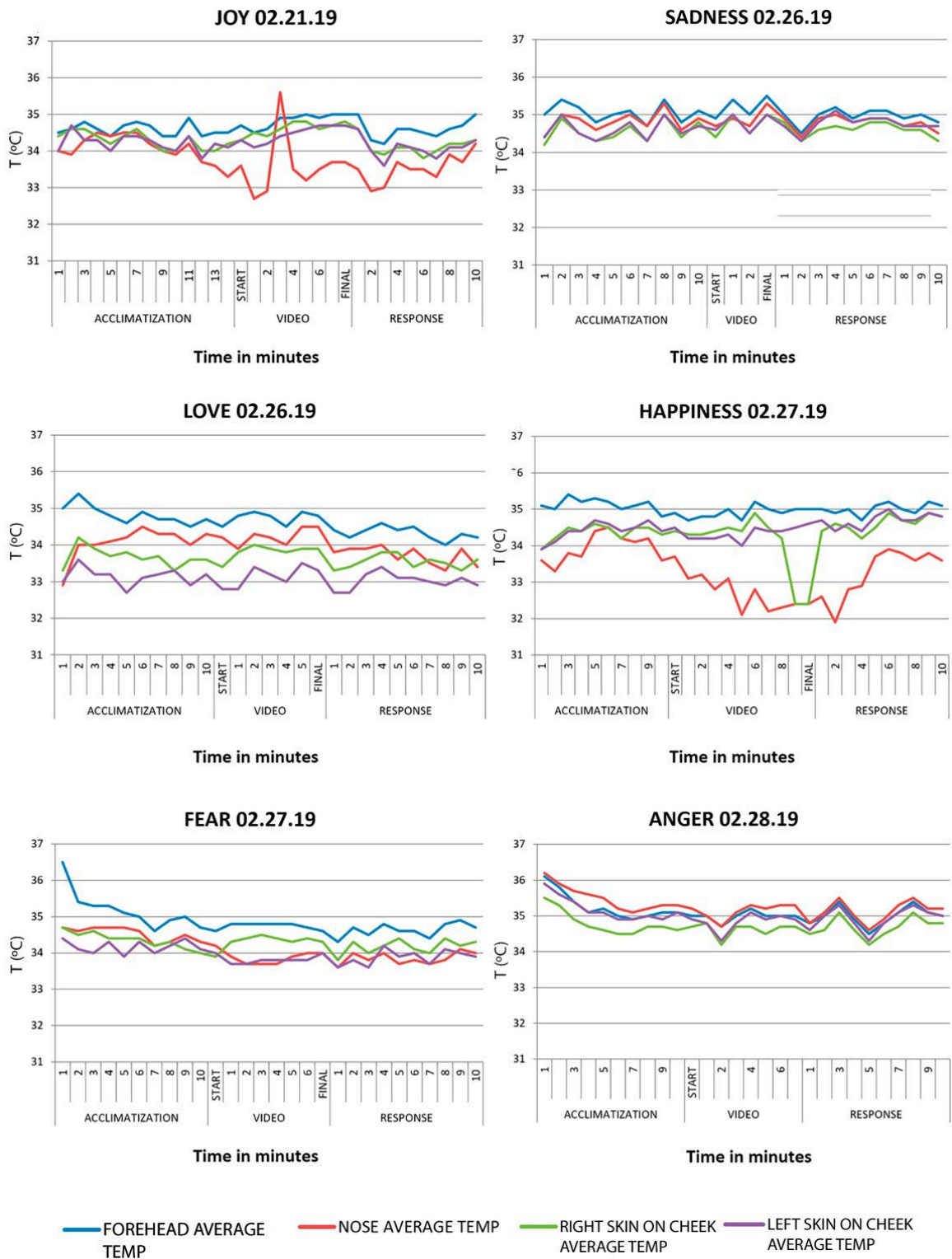

**Figure 5.** Lineal graphs showing the variation of the average temperature by facial areas during part A of the study.

Table 4. Facial average temperatures in each phase for the evaluated Regions of Interest (ROIs).

| Joy | Acclimatization | | Video | | Response | |
|---|---|---|---|---|---|---|
| | Start | Final | Start | Final | Start | Final |
| Forehead | 34.5 | 34.5 | 34.7 | 35.0 | 35.0 | 35.0 |
| Nose | 34.0 | 33.3 | 33.6 | 33.7 | 33.5 | 34.2 |
| Right skin on cheek | 34.4 | 34.2 | 34.3 | 34.8 | 34.6 | 34.3 |
| Left skin on cheek | 34.0 | 34.1 | 34.3 | 34.7 | 34.6 | 34.3 |
| **Sadness** | **Acclimatization** | | **Video** | | **Response** | |
| | Start | Final | Start | Final | Start | Final |
| Forehead | 35.0 | 35.1 | 34.9 | 35.5 | 35.0. | 34.8 |
| Nose | 34.4 | 34.9 | 34.7 | 35.3 | 34.9 | 34.5 |
| Right skin on cheek | 34.2 | 34.8 | 34.4 | 35.0 | 34.8 | 34.3 |
| Left skin on cheek | 34.4 | 34.7 | 34.6 | 35.0 | 34.7 | 34.7 |
| **Love** | **Acclimatization** | | **Video** | | **Response** | |
| | Start | Final | Start | Final | Start | Final |
| Forehead | 35.0 | 34.7 | 34.5 | 34.8 | 34.4 | 34.2 |
| Nose | 32.9 | 34.3 | 34.2 | 34.5 | 33.8 | 33.4 |
| Right skin on cheek | 33.3 | 33.6 | 33.4 | 33.9 | 33.3 | 33.6 |
| Left skin on cheek | 33.0 | 33.2 | 32.8 | 33.3 | 32.7 | 32.9 |
| **Happiness** | **Acclimatization** | | **Video** | | **Response** | |
| | Start | Final | Start | Final | Start | Final |
| Forehead | 35.1 | 34.8 | 34.9 | 35.0 | 35.0 | 35.1 |
| Nose | 33.6 | 33.6 | 33.7 | 32.4 | 32.6 | 33.6 |
| Right skin on cheek | 33.9 | 34.3 | 34.4 | 32.4 | 34.4 | 34.8 |
| Left skin on cheek | 33.9 | 34.4 | 34.5 | 34.6 | 34.7 | 34.8 |
| **Fear** | **Acclimatization** | | **Video** | | **Response** | |
| | Start | Final | Start | Final | Start | Final |
| Forehead | 36.5 | 34.7 | 34.6 | 34.6 | 34.3 | 34.7 |
| Nose | 34.7 | 34.3 | 34.2 | 34.0 | 33.6 | 34.0 |
| Right skin on cheek | 34.7 | 34.0 | 33.9 | 34.3 | 33.8 | 34.3 |
| Left skin on cheek | 34.4 | 34.1 | 34.0 | 34.0 | 33.6 | 33.9 |
| **Anger** | **Acclimatization** | | **Video** | | **Response** | |
| | Start | Final | Start | Final | Start | Final |
| Forehead | 36.1 | 35.1 | 35.0 | 35.0 | 34.8 | 35.0 |
| Nose | 36.2 | 35.3 | 35.2 | 35.3 | 34.8 | 35.2 |
| Right skin on cheek | 35.5 | 34.6 | 34.7 | 34.7 | 34.5 | 34.8 |
| Left skin on cheek | 35.9 | 35.1 | 34.9 | 34.9 | 34.6 | 35.0 |

*3.2. Part B: Thermal Changes with Audio Stimulus (Music)*

Acclimatization phase. A generalized decrease in temperature was observed in all of the emotions observed followed by an increase in temperature over a period of a few minutes.

Stimulus phase. There was a constant increase in temperature in all of the emotions except in anger, which descended slightly. In the emotion fear, at the end of the stimulus, a decline in the temperature is confirmed. Once again, we were able to observe different values between the nose and the other facial areas being studied.

Response phase. The results showed an increase in temperature in the emotions joy, fear, anger and sadness. In love and happiness, the facial temperature went down due to the body returning to its basal temperature.

Table 5 provides details of thermal change undergone in the facial areas chosen in this study during the beginning and end of each phase (acclimatization, video, and response), which was registered in the subject during the audio stimulus (music) of the emotions chosen.

Table 5. Facial average temperatures in each phase.

| Happiness | Acclimatization | | Music | | Response | |
|---|---|---|---|---|---|---|
| | Start | Final | Start | Final | Start | Final |
| Forehead | 34.6 | 34.2 | 33.4 | 33.9 | 33.2 | 33.4 |
| Nose | 30.6 | 30.4 | 30.1 | 29.3 | 28.4 | 29.0 |
| Right skin on cheek | 32.0 | 31.6 | 31.4 | 32.3 | 31.7 | 31.9 |
| Left skin on cheek | 31.8 | 31.2 | 31.4 | 31.7 | 31.2 | 30.9 |
| **Joy** | **Acclimatization** | | **Music** | | **Response** | |
| | Start | Final | Start | Final | Start | Final |
| Forehead | 34.0 | 34.0 | 33.9 | 34.0 | 34.0 | 33.9 |
| Nose | 27.4 | 27.6 | 28.3 | 28.7 | 28.3 | 29.7 |
| Right skin on cheek | 30.7 | 30.5 | 31.1 | 31.4 | 31.7 | 31.3 |
| Left skin on cheek | 31.0 | 30.2 | 30.7 | 31.0 | 31.4 | 31.3 |
| **Sadness** | **Acclimatization** | | **Music** | | **Response** | |
| | Start | Final | Start | Final | Start | Final |
| Forehead | 34.1 | 33.9 | 33.7 | 34.0 | 34.2 | 33.7 |
| Nose | 29.4 | 30.9 | 30.2 | 31.3 | 32.0 | 31.0 |
| Right skin on cheek | 32.5 | 32.5 | 32.4 | 33.0 | 33.1 | 32.5 |
| Left skin on cheek | 32.1 | 32.0 | 32.1 | 32.5 | 32.8 | 32.5 |
| **Fear** | **Acclimatization** | | **Music** | | **Response** | |
| | Start | Final | Start | Final | Start | Final |
| Forehead | 34.3 | 34.2 | 33.9 | 34.1 | 33.4 | 34.2 |
| Nose | 28.7 | 29.1 | 28.8 | 28.9 | 29.3 | 30.0 |
| Right skin on cheek | 31.4 | 31.2 | 31.2 | 31.5 | 31.2 | 32.1 |
| Left skin on cheek | 31.4 | 31.3 | 31.2 | 31.2 | 30.5 | 31.9 |
| **Anger** | **Acclimatization** | | **Music** | | **Response** | |
| | Start | Final | Start | Final | Start | Final |
| Forehead | 33.8 | 33.8 | 33.6 | 33.4 | 33.4 | 33.8 |
| Nose | 29.3 | 30.5 | 30.7 | 28.5 | 29.4 | 29.8 |
| Right skin on cheek | 31.6 | 32.0 | 31.7 | 32.0 | 31.9 | 32.4 |
| Left skin on cheek | 31.4 | 32.1 | 31.6 | 31.5 | 31.4 | 32.1 |
| **Love** | **Acclimatization** | | **Music** | | **Response** | |
| | Start | Final | Start | Final | Start | Final |
| Forehead | 34.7 | 34.1 | 34.3 | 34.2 | 34.2 | 33.3 |
| Nose | 29.7 | 30.0 | 30.1 | 30.7 | 30.9 | 29.8 |
| Right skin on cheek | 33.2 | 33.1 | 33.3 | 33.6 | 33.2 | 32.1 |
| Left skin on cheek | 33.5 | 33.2 | 33.2 | 33.3 | 33.3 | 32.3 |

The tendency of the average temperature explained during the audio stimulus in the study of the case can be observed in Figure 6.

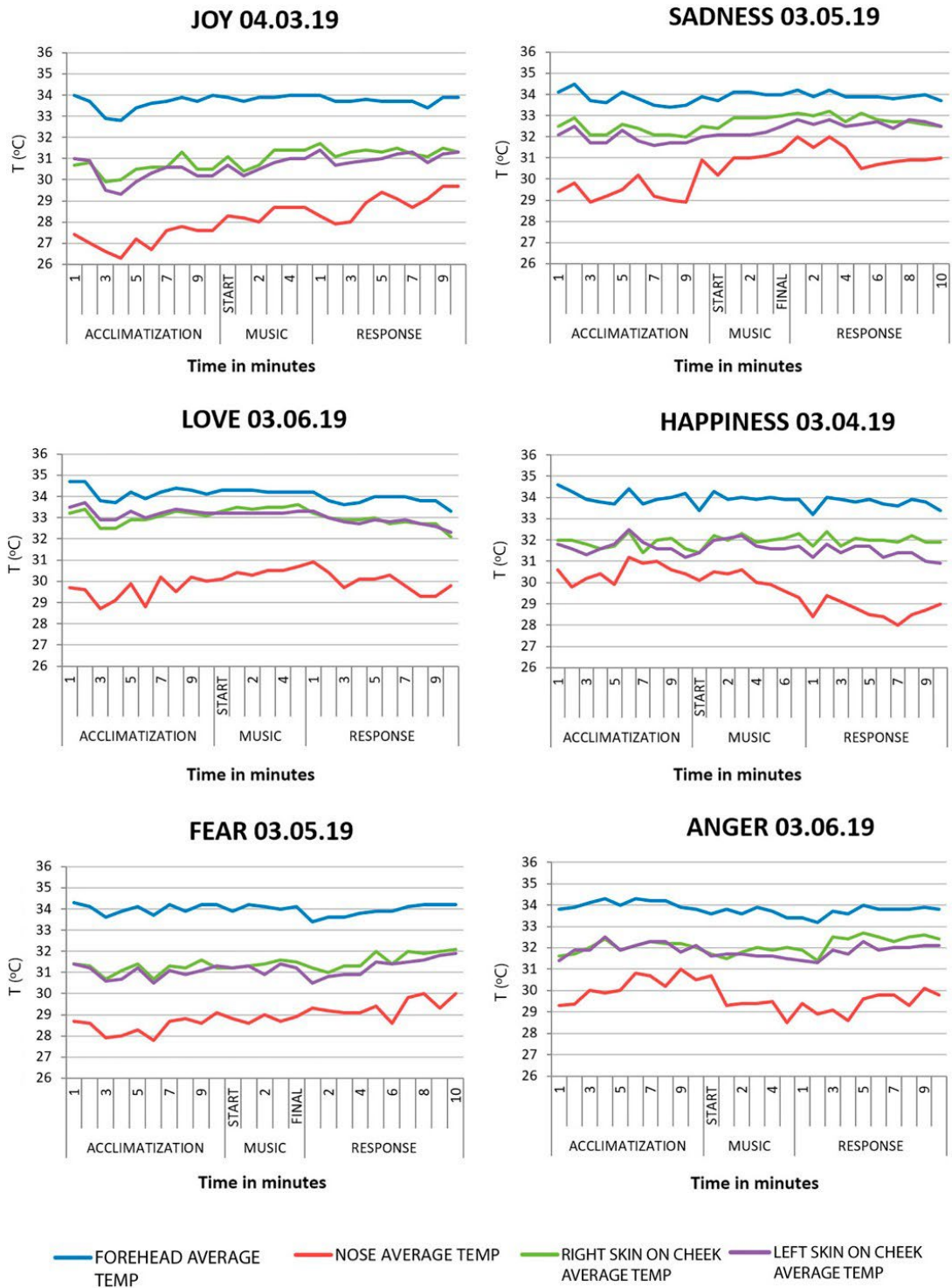

**Figure 6.** Lineal graphs showing the variation of the average temperature by facial areas during part B of the study.

For a better understanding of the results, Table 6 details the differences and similarities in each step of each emotion, and under both the video and musical stimulus.

Table 6. Comparison of results of parts A and B of the study.

|  |  | Temperature | |
|---|---|---|---|
|  |  | **Video** | **Music** |
| **Love** | Acclimatization | Increase and decrease | Decrease |
|  | Stimulus | Increase | Increase |
|  | Response | Decrease | Decrease |
|  | Average Graph | Increase and decrease, increase in stimulu | Increase |
| **Happiness** | Acclimatization | Increase and decrease | Decrease |
|  | Stimulus | Decrease | Increase |
|  | Response | Increase | Increase and decrease |
|  | Average Graph | Decrease in stimulation: nose, right skin on cheek. Increase in stimulation: forehead, left skin on cheek | Decreases in stimulus in the nose, slight increase in the rest |
| **Fear** | Acclimatization | Decrease | Decrease |
|  | Stimulus | Increase | Increase |
|  | Response | Increase | Increase |
|  | Average Graph | Decrease, increase during stimulus | Slight and gradual increase |
| **Anger** | Acclimatization | Decrease | Increase |
|  | Stimulus | Decrease and increase | Decrease |
|  | Response | Increase | Increase |
|  | Average Graph | Increase in stimulus | Slight decrease in stimulus |
| **Joy** | Acclimatization | Decrease and increase | Decrease |
|  | Stimulus | Increase | Increase |
|  | Response | Increase | Increase |
|  | Average Graph | Acute increase in stimulus | Increase |
| **Sadness** | Acclimatization | Increase | Decrease and increase |
|  | Stimulus | Increase | Increase |
|  | Response | Decrease | Decrease |
|  | Average Graph | Decrease and increase | Decrease and increase |

*3.3. Part C: Sequential Thermographic Evaluation*

The thermographic images obtained would allow the students to advance in the acquisition of emotional skills, given that they themselves are able to evaluate the results. This learning is facilitated even more through the visualization of the temporal sequence images when faced with a stimulus. For example, the temporal sequence of the emotion rage, shown in Figure 7, when subjected to an audiovisual stimulus. The color in the images represents the temperature (purple, lower temperature and yellow; higher temperature).

In the stimulus phase, a gradual increase in the temperature in general can be observed, the cheeks being the most notable. Figure 7a–c show the thermal changes in the face when the subject is exposed to an audiovisual stimulus. We can see a significant change between Figure 7a,b at the beginning and or minutes into the stimulus, respectively. Facial temperature increases as its subject receives the stimulus, and is maintained constantly until it is finished (Figure 7c).

In the response phase (Figure 7d–f), we can see how the temperature of the nose decreases as does that of the forehead. Of the facial areas, it is the cheeks that take longer to return to the normal temperature. Note that this type of qualitative evaluation which allows a very rapid appreciation of the stimulus–response relationship, should always be supported by quantitative values such as those collated in the previous tables and figures.

For the participating student, it was very easy for he or she to read the sequence of the images.

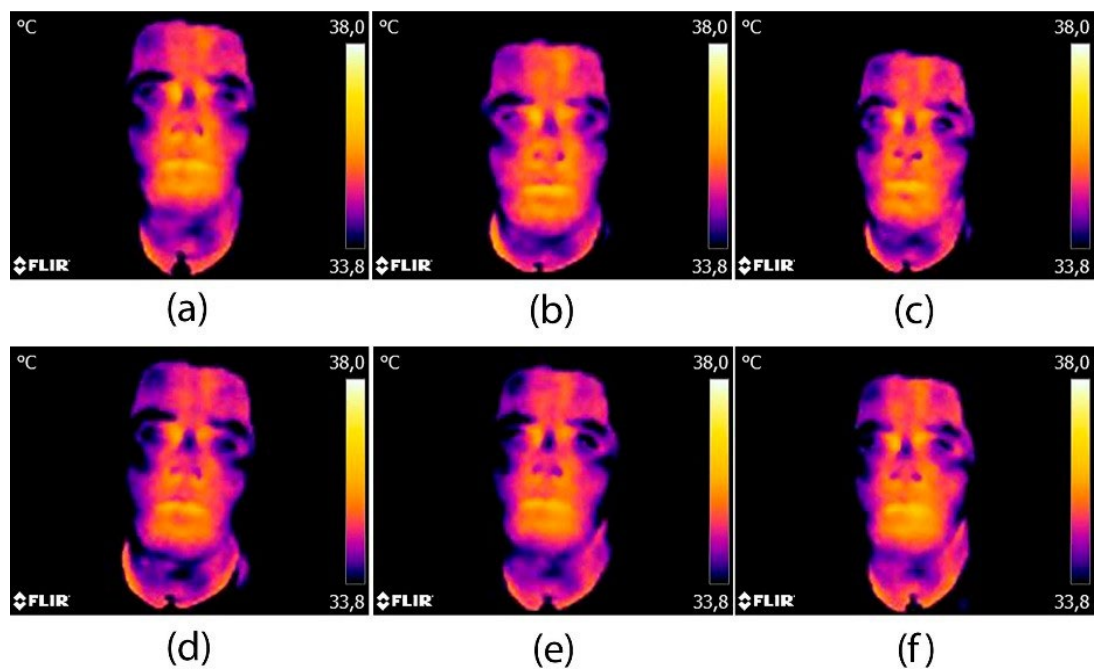

**Figure 7.** Example of a temporal series of thermographic images for the anger emotion: (**a**) beginning of stimulus; (**b**) 4 min of stimulus; (**c**) end of stimulus and beginning of the response period; (**d**) one minute after the stimulus; (**e**) six minutes after the stimulus; (**f**) ten minutes after the stimulus (end of recovery period).

## 4. Discussion

The motivation for this study arose from the interest in developing emotional skills in nursing students by means of simulation techniques. The literature has highlighted that the emotional skills are fundamental in the nursing profession [33], and that the emotional intelligence of the health professionals have an influence on clinical decision taking, well-being in workgroups, and in performance [34]. Even so, it has been detected that emotional skills are often forgotten in the nursing curriculum [3] so the context offers an opportunity for our research.

There are many studies into the skills related to the learning-teaching processes. However, more research is needed into which are the best strategies to achieve, apply, and evaluate each of the skills [5]. Previous studies have evaluated the emotional states in simulation of critical patients through questionnaires pre- and post-operation [18]. However, in our case, our interest is in applying a more quantifiable methodology which is easy to apply in order to monitor the emotional behavior of the students and with the possibility of improving the learning-teaching processes. Infrared thermal imaging has been considered as a promising methodology in the emotional context, given that it is a non-invasive technique [35]. Therefore, this technique brings together all of our expectations in identifying the emotions by means of the thermographic camera.

This study has been carried out with the case methodology, taking thermographic images in which a single subject was exposed to stimuli from two sources: video and music. The emotions studied were: Joy, Love, Happiness, Sadness, Fear and Anger, and the temperature was measured on the forehead, nose, and left and right cheeks. The results are collated on the basis of the acclimatization, stimulus, and response phase.

Our study applied the video to generate emotions by watching certain films. We have not come across studies using the same intervention, but we have, however, come across similar research carried out with medical students, with the objective of improving the training of palliative care [36].

In relation to the parameters measured during the acclimatization period in both stimuli, it has been observed that the temperature did not follow a pattern of increasing or decreasing. This could give rise to two questions. In the first place, a large thermal change experienced by the subject on coming from the exterior of the building to the room in which the experiment was to take place was much colder outside than it was in the room. In the second place, it was not possible to keep the temperature constant within the room, an aspect that was considered in the previous studies in which the temperature was kept constant between 18.0 °C–18.3 °C [35,37,38], or in 22 °C.

Regarding the thermal camera (FLIR 6), it was used for this first approach to assess emotional competencies in nursing students for three main reasons. First of all, we wanted a low-cost approach; therefore, it can be easily extrapolated to other faculties since it is easily acquired by universities. Secondly, despite being a low-definition camera, it complies with the study approach, since it allows for extracting the significant statistical values of the defined ROI (mainly the mean facial temperature) to extract emotional patterns. Lastly, with respect to the subject–camera distance, the distance is higher than the minimum focus distance (0.5 m) to prevent the camera from disturbing the subject if it was too close to the face of the subject.

Additionally, in relation to the thermal device, using a higher accuracy thermal camera can lead to more accurate temperature measurements but will increase the deployment costs [39]. As an emotion induces a gradual change according to the baseline stablished in the acclimatization phase, we aimed to analyzed the global trend (increase/decrease) rather than the fluctuations (which are considered not significant due to the aforementioned camera precision). Despite the camera's absolute error, the ROI results allowed to identify different facial patterns for the emotions, which are adequate for the acquisition and evaluation of emotional competences for Nursing degree students. Therefore, for our study, which is closer to a classification problem, the accuracy of temperature measurement using IR camera did not influence significantly the study as it could have done in a regression problem. Regarding the use of low precision camera, authors would like to add that, in Li et al. [40] and Aryal and Becerik-Gerber [39], a very low-cost Flir Lepton camera with an accuracy of ±5 °C and 80 × 60 pixels was used for ROI temperature extraction, and authors showed that the camera was adequate for monitoring changes in skin temperature for thermal comfort assessment.

In the stimulus period, the findings reflect a thermal change in the facial area of the subject with an ascending tendency with the video and music. The results are in agreement with the studies of Ioannuou [35], who demonstrated that the subject who was subjected to the stimulus experienced a dilation of the vessels in the facial regions.

Anger had a slightly different behavior, given that, during the stimulus with music, the temperature went down. With the exception of anger, our findings differ from the results obtained by Goulart [32], who indicated a decrease in facial temperature when experiencing emotions considered negative (anger, sadness and fear). This could be because our study was an isolated case which might have obtained other results with a greater population.

The findings in the response period showed an increase and decrease the emotions. No previous studies have been found in which changes in facial temperature in the response period have been detected. There is a considerable lack of literature with which to discuss these findings. However, and given that we are carrying out the study with just one case, we can give a brief justification. The temperatures increase in joy, anger, and fear, possibly because the subject continues to experience the sensations created by the stimulus. In love, happiness, and sadness, it might have gone down given that the temperature of the room went down 1 °C during the study.

Comparing the results obtained during the video and music stimulus, a different change of temperature in the nose was observed for each stimulus. Goulart also observed changes in perinasal fluctuations [32]. The result may be justified given that it is a distal facial area and experiences different thermal changes than the rest of the face. However, Cruz-Albarran observed different changes in the temperature of the nose for each stimulus, but with a pattern similar to the rest of the facial areas measured [29].

In almost all of the emotions, the facial temperature pattern has remained similar in all the periods. The explanation may be due to an inadequate choice of stimulus, that is, the type of music or films chosen. For this reason, we may have used a bank of images like the Affective Picture System (IAPS), used by [41]. Another explanation could be that the subject of this study is the same person involved in its drawing up, which could have conditioned the response of his or her emotions on knowing what the stimuli were. In this respect, Kosonogov et al. suggested using subjects not involved in the research [38,42].

This study was a first approximation to apply thermography to the evaluation of the emotions through changes in facial temperature. The study highlights that it is necessary to include other areas, and that complexity for that may be found in the laboratories themselves [12]. In our case, we have made good on this question by using these musical and visual stimuli, avoiding the intervention of other people and including other facial areas measured.

It could be that changes in facial temperature do not give responses to all that is demanded in the evaluation and monitoring of the emotions. Given that the facial muscles play a fundamental role in the expression of emotions, thermography might make up for some of these shortcomings in the said evaluation [12].

Our findings provide evidence on the use of the Thermography technique in the monitoring of the emotions of nursing students. The objective of this case is to encourage the education community to experiment with the technique, bring it into the emotional skills curriculum, and use it to create a bank of case studies useful for students and teachers.

*4.1. Conclusions*

This study has carried out a first approximation to propose thermography as a teaching technique for the evaluation of the emotions, and as best material for virtual laboratories. Derived from this study, the following can be concluded:

1. Whenever the student is exposed to a stimulus, there is a thermal bodily response that demonstrates that there is a gradient thermal change with respect to the basal temperature of the subject.
2. All of the facial areas follow a common thermal pattern in response to the stimulus, with the exception of the nose.
3. During the acclimatization phase, the body temperature of the subject did not follow a standard pattern. The room in which the study was carried out varied considerably in temperature throughout the sessions which meant that the temperature of the subject was not always measured under the same conditions.
4. It is recommended that the subject of the study is not involved in the research into the case in order to not condition their thermal responses.
5. Thermography is the techniques suitable for simulation practices in emotional skills given that it is non-invasive, it is quantifiable, and easy to access.

*4.2. Limitations and Future Lines of Research*

The main limitation of the study is the size of the sample, limiting the study to just one case. A larger sample would allow us to quantify the optimal response and stimulus times for each emotion. At the same time, the subject being studied is part of the study itself, which limited the obtaining of the results to a large degree. Another of the limitations detected in this work is the difficulty in generating audiovisual stimuli for the whole range of emotions that will have to be managed by future nurses.

Another limitation of this study would be the main sources of error identified in the study; firstly, the camera specifications, since a device with better temperature accuracy and spatial resolution will allow for measuring the temperature of the human face with its irregularities. However, the employed low-cost camera can be considered suitable for the research aim (competence learning). Secondly, the ambient parameters, namely room temperature and humidity. Although during the sessions it could be considered stable, there were slight differences, since it was not possible to keep constant. Therefore, if the ambient parameters could be kept constant, the temperature trends due to the emotions could be identified more clearly. In addition, the ROIs definition could be increased in number for a more complete analysis. It was not possible due to the camera's spatial resolution. However, the four selected ROIs allowed for identifying the emotional changes adequately.

Future lines of investigation will look at the application of thermography in the training of emotional skills in nursing students. It will also go into the drawing up of teaching material banks of virtual laboratories focused on managing emotional skills to a greater depth. All of this is oriented at achieving a better academic performance.

**Author Contributions:** Conceptualization, P.M.-S. and P.R.-G.; methodology, P.R.-G. and J.A.B.-A.; software, P.R.-G.; validation, R.G.-G., L.Á.-B., and C.L.-P.; formal analysis, P.R.-G. and R.G.-G.; investigation, P.M.-S. and J.A.B.-A.; resources, L.Á.-B.; data curation, R.G.-G.; writing—original draft preparation, P.M.-S.; writing—review and editing, C.L.-P.; visualization, J.A.B.-A.; supervision, P.R.-G. All authors have read and agreed to the published version of the manuscript.

**Funding:** This research received no external funding.

**Conflicts of Interest:** The authors declare no conflict of interest.